\documentclass[preprint,showpacs,preprintnumbers,amsmath,amssymb]{revtex4}
\usepackage{graphicx}
\usepackage{dcolumn}
\usepackage{bm}
\def\slash#1{\not\!#1}

\def\delsla{\!\!\not\!\partial}

\begin{document}
\preprint{OCU-PHYS-296}

\title{Phase diagram of chiral and diquark condensates \\
at finite temperature and density \\
in the 2-dimensional Gross Neveu model}

\author{Hiroaki Kohyama}
 \email{kohyama@sci.osaka-cu.ac.jp}
\affiliation{%
Department of Physics, Osaka City University,
Sumiyoshi-ku, Osaka 558-8585, JAPAN
}%

\date{\today}

\begin{abstract}
We construct the phase diagram of the chiral and diquark condensates
at finite temperature and density in the $1+1$ dimensional (2D) two flavor massless
Gross Neveu model. The resultant phase diagram shows (I) the chiral
condensed phase at low temperature and density, (II) the diquark condensed phase at
low temperature and high density, and (III) the chiral and diquark coexisting phase
at low temperature and intermediate density. This phase structure is also seen in
the 3D Gross Neveu model and the 4D Nambu Jona-Lasinio (NJL) model. Thus the phase
diagrams of the chiral and diquark condensates in the NJL-type models do not
change qualitatively in 2D, 3D and 4D.
\end{abstract}

\pacs{12.38.Aw, 12.38Lg, 11.15.Pg, 11.10.Wx}
\maketitle
\section{\label{sec:level1}Introduction}

Studying the phase structure of Quantum Chromodynamics (QCD) is one of the most
interesting topics in the field of strong interaction physics.
QCD is an asymptotically free theory \cite{Gross}, and the interaction between
quarks and gluons becomes weak at high temperature and density. Then the quarks
can be free particles at high temperature and density, and this system
is called the quark-gluon-plasma. On the other hand, at low temperature and density,
quarks are confined into hadrons and can not be observed as free particles.
Furthermore, it is widely believed that
the color superconducting phases are realized at low temperature and
high baryon density. The color superconductivity is the state where the
quark-quark (diquark) Cooper pairs \cite{Cooper} become correlated, which condense.
Since it was indicated that color superconducting phases have non-negligible region
in the QCD phase diagram\cite{Wilczek}, there has been a lot of works on it.
For review articles, see \cite{colorsuper}. Color superconductivity is important
because it might be observed in compact stars, such as neutron stars.
The density of a neutron star is about ten times of that
of the normal nuclear. This density corresponds to a quark chemical potential
$\mu \sim 500{\rm MeV}$ where the deconfinement phase transition may
happen. Thus it is physically reasonable to expect that color superconducting
states are realized in neutron stars.

Investigation of these subjects from QCD is still a challenge to theorists, and we
usually have to rely on some effective theories of QCD. One of the most important
models is the Nambu Jona-Lasinio (NJL) model \cite{NJL} which is a low energy effective
theory of QCD. Since the NJL model successfully describes the phenomena of the
transition from chiral broken phase to chiral restored phase, a variety
of works has been devoted to the study based on it \cite{onNJL}.
Recently, it was found that the NJL model also describes the phase transition from
chiral broken phase to color superconducting phase \cite{Mei}, and the phase diagram
of the NJL model shows similar structure with the QCD phase diagram.
Thus the NJL model is also effective when we consider a color superconductivity.

The study of the phase structure of the NJL model in lower dimensions (D) is also
an interesting issue, because the model usually becomes simpler in lower dimensions
\cite{Weinberg}. Indeed, the NJL model is renormalizable in 2D and 3D. The NJL-type
model in lower dimensions is the Gross Neveu (GN) model, which is proposed in
1974 \cite{GN}. The GN model shares many properties with QCD, such as asymptotic
freedom and chiral symmetry breaking in vacuum, and there has been a lot of works
on it \cite{Rosen,onGN,Ulli,Klimenko,Kohyama,Zhou}. The phase diagram of 
quark-antiquark condensate
in the 2D and 3D GN model was obtained in \cite{Ulli} and \cite{Klimenko}. By using
the mean-field approximations, the phase structure of the quark-antiquark ($q\bar{q}$)
and diquark ($qq$) condensates in the 3D GN model with 4-component spinor quarks was
derived in \cite{Kohyama}. Although the possibility of spontaneously symmetry breaking
is excluded by the Mermin-Wagner-Coleman theorem, however, I believe that the study
of the $q\bar{q}$ and $qq$ condensates near the Fermi surface is important from the
phenomenological point of view. Actually, it was shown that the $q\bar{q}$ and $qq$
condensates could be formed in the 3D GN model by using the effective potential
analysis \cite{Kohyama}. The resultant phase diagrams show
considerably similar structure to the case of the 4D NJL model. This may suggest
that the phase structure of the quark matter does not change qualitatively in
the cases of lower dimensions. To confirm this idea, we focus on the phase
structure in the 2D GN model here. Recently, within the 2D GN model, the $q\bar{q}$
and $qq$ condensates in vacuum (zero temperature and chemical potential) has been
studied in \cite{Zhou}, and it has been found that the competition between
these condensates is sensitive to the ratio of the coupling strengths of the
$q\bar{q}$ and $qq$ condensates.

In this paper, we study the $q\bar{q}$ and $qq$ condensates at finite temperature
and density in the 2D two flavor massless GN model, and obtain the phase diagrams.
Then we compare the result with those of the 3D GN model and the 4D NJL model.

The paper is constituted as follows: In Sec.II, we introduce the Lagrangian
density of the 2D GN model and apply the mean-field approximation.
In Sec.III, the thermodynamic potential is obtained.
Then we display the numerical results for the $q\bar{q}$ and $qq$ condensates
at zero and finite temperatures in Sec.IV. The phase diagrams of the 2D GN model
are shown in Sec.V. Finally in Sec.VI, we compare the results
of the 2D, 3D GN models and 4D NJL model, and discuss the
similarities and differences of them.

\section{2D Gross Neveu Model}

\subsection{Preliminary}
We consider the following Lagrangian density as the original form of the
four fermion interaction model,
\begin{align}                                                                 
\mathcal{L} = \bar{q}i \delsla q + G_S(\bar{q}q)^2
\label{oGN}.
\end{align}                                                                   
Here $G_S$ is the coupling constant of the $q\bar{q}$ condensate. To incorporate
the diquark condensate in this Lagrangian, we apply the Fiertz transformation to
$(q\bar{q})^2$ term and leave the most relevant interaction near the Fermi
surface \cite{Wilczek}. More concretely, the diquark interacting term
is determined as follows: Let us write the diquark term as
$(q^{\rm T} \mathcal{O} q)^2$ in its most general form, where $\mathcal{O}$ is an
operator in spinor, flavor and color spaces. The Pauli principle,
\begin{align}                                                                 
q^{\rm T} \mathcal{O} q = \mathcal{O}_{ij} q_i q_j
= -\mathcal{O}_{i j} q_j q_i = - q^{\rm T} \mathcal{O}^{\rm T} q,
\end{align}                                                                   
indicates that $\mathcal{O}$ is a totally antisymmetric operator. Since the
attractive interaction becomes the dominant contribution near the Fermi surface
\cite{Wilczek}, we select the color antitriplet matrices
($\lambda_2, \lambda_5, \lambda_7$) in color space. The renormalization group
arguments \cite{Son:1998uk} force the operator in spinor space to be antisymmetric.
Then the operator in flavor space has to be antisymmetric. Observing the above
mentioned consideration, we choose the Lagrangian density of the form
\begin{align}                                                                 
\mathcal{L} &=  \bar{q}i \delsla q
   + G_S (\bar{q}q)^2
   + G_D \sum_{a=2,5,7} (\bar{q} i \tau_2 \lambda_a \gamma^5 q^{C})
      (\bar{q}^{C} i \tau_2 \lambda_a \gamma^5 q),
\label{pGN}
\end{align}                                                                   
where $\tau_2$ is the second Pauli matrix in flavor space and $C$ in $q^C$ expresses
the charge conjugation. The charge conjugated fields are given by
$q^C = C\bar{q}^{\rm T}, \, \, \bar{q}^C=q^{\rm T} C$, with the charge conjugation
matrix $C$. Since a vector in color space can be rotated into one direction
by a global $SU(3)$-color transformation, one can select a color direction to
{\it blue} without loss of
generality. This is equivalent to leave $\lambda_2$ in Eq.(\ref{pGN}), and we have the
diquark term
$(\bar{q} i \tau_2 \lambda_2 \gamma^5 q^{C})(\bar{q}^{C} i \tau_2 \lambda_2 \gamma^5 q)$.

\subsection{The model}
Following the reason discussed in the previous subsection, we employ the Lagrangian
density for the 2D two flavor massless GN model with the diquark condensate term,
\begin{align}                                                                 
\mathcal{L} = \bar{q}i \delsla q + G_S (\bar{q}q)^2
    + G_D (\bar{q} i \tau_2 \lambda_2 \gamma^5 q^C)
          (\bar{q}^C i \tau_2 \lambda_2 \gamma^5 q).
\label{GN}
\end{align}                                                                   
In 2D, $q \equiv q_{\alpha k}(t,x)$ is two component spinor fields with the two flavors
$\alpha= u,\,d$ and three colors $k=${\it r} (red), {\it g} (green), {\it b} (blue).
For $\gamma$ matrix, we use the following representation,
\begin{align}                                                                 
\gamma^0 = 
\left(
\begin{array}{cc}
1 & 0\\
0 & -1
\end{array}\right)\;,
\gamma^1 = 
\left(
\begin{array}{cc}
0 & 1\\
-1 & 0
\end{array}\right) = -C,
\quad \gamma^5 = \gamma^0 \gamma^1.
\end{align}                                                                   

The Lagrangian Eq.(\ref{GN}) is invariant under the following transformations:
\begin{quote}
1. Parity $\mathcal{P}$: $q(t,x) \rightarrow \gamma^0 q(t,-x)$ \\
2. Time reversal $\mathcal{T}$:
   $q(t,x) \rightarrow \gamma^0 q(-t,x)$ \\
3. Charge conjugation $\mathcal{C}$: $q^C \rightarrow 
   -\gamma^1 \bar{q}^{\rm T},\quad
   \bar{q}^C \rightarrow -q^{\rm T} \gamma^1$ \\
4. Discrete chiral symmetry $\mathcal{X}_D$: $q \rightarrow \gamma^5 q$.
\end{quote}
The symmetry properties are discussed in the paper \cite{Zhou}
in more detail.

Within the mean-field approximation, the Lagrangian density turns out to be
\begin{align}
\tilde{\mathcal{L}} = \bar{q}i \delsla q 
     -\bar{q} \sigma q 
    &{}+ \frac{1}{2}\Delta^{*}(\bar{q}^C i \tau_2 \lambda_2 \gamma^5 q)
    + \frac{1}{2}\Delta(\bar{q} i \tau_2 \lambda_2 \gamma^5  q^C)
    -\frac{\sigma^2}{4 G_S} - \frac{|\Delta|^2}{4 G_D}
\label{mGN}.
\end{align}
Here $\sigma$ and $\Delta$ are the order parameters for the chiral and diquark condensates,
\begin{eqnarray}
    \sigma = -2G_S \langle \bar{q}q \rangle 
    \quad {\rm and} \quad
    \Delta = 2G_D \langle \bar{q}^C i \tau_2 \lambda_2 \gamma^5  q \rangle.
\end{eqnarray}

\section{The thermodynamic potential}
\subsection{Derivation of the thermodynamic potential}
The partition function is evaluated by using the standard procedure,
\begin{align}                                                                 
\mathcal{Z} = N^{\prime} \int [d\bar{q}][dq] \exp
  \biggl\{
    \int_0^{\beta}\!\!\! d\tau \! \int \!\! dx 
       \bigl( \tilde{\mathcal{L}}+\mu \bar{q}\gamma_0 q \bigr)
  \biggr\},
\end{align}                                                                   
where $\beta=1/T$ is the inverse temperature and $\mu$ is the quark chemical potential.
Introduction of the Nambu-Gorkov basis \cite{Nambu}
\begin{align}                                                                 
     \Psi = 
     \left(
     \begin{array}{c}
     q \\
     q^C
     \end{array}\right)
     \quad {\rm and} \quad
     \bar{\Psi} = 
     \left(
     \, \bar{q} \,\,\, \bar{q}^C
     \right) \, ,
\end{align}                                                                   
enables us to write the partition function in the simple form,
\begin{align}                                                                 
\mathcal{Z} = & N^{\prime} \exp
  \biggl\{
    - \int_0^{\beta}\!\!\! d\tau \! \int \!\! dx 
       \biggl( \frac{\sigma^2}{4G_S}+\frac{|\Delta|^2}{4G_D} \biggr)
  \biggr\} \nonumber \\
  & \times 
  \int [d\Psi] \exp
  \biggl\{
    \frac{1}{2} \sum_{n,\,p} \bar{\Psi} \bigl( \beta G^{-1} \bigr) \Psi
  \biggr\}.
\end{align}                                                                   
The matrix $G^{-1}$ is defined by
\begin{align}                                                                 
  G^{-1} =
  & \left(
    \begin{array}{cc}
       (\slash{p} - \sigma + \mu \gamma^0){\bf 1}_f {\bf 1}_c
     & i \tau_2 \lambda_2 \gamma^5 \Delta \\
       i \tau_2 \lambda_2 \gamma^5 \Delta^{*}
       & (\slash{p} - \sigma - \mu \gamma^0){\bf 1}_f {\bf 1}_c
    \end{array}\right),
\end{align}                                                                   
where ${\bf 1}_f$ and ${\bf 1}_c$ are the unit matrices in flavor and color
spaces, respectively.

Then the thermodynamic potential $\Omega = -\ln \mathcal{Z} / \beta V$ becomes
\begin{align}                                                                 
     \Omega  (\sigma,|\Delta|) = 
       \frac{\sigma^2}{4G_S} + \frac{|\Delta|^2}{4G_D}
       - \frac{1}{\beta V} \ln \int [d\Psi] \exp 
     \left[ 
     \frac{1}{2}\sum_{n,\,p} \bar{\Psi} (\beta G^{-1}) \Psi 
     \right],
\end{align}                                                                   
where $V$ is the volume of the thermal system. With the help of the formula
\begin{align}                                                                 
     \int [d\Psi] \exp 
     \left[ 
     \frac{1}{2}\sum_{n,\,p} \bar{\Psi} (\beta G^{-1}) \Psi 
     \right]
     =
     {\rm det}^{1/2} (\beta G^{-1}) \, ,
\end{align}                                                                   
we have
\begin{align}                                                                 
     \Omega & (\sigma,|\Delta|) = 
      \frac{\sigma^2}{4G_S} + \frac{|\Delta|^2}{4G_D}
       - \frac{1}{\beta V} \ln {\rm det}^{1/2} (\beta G^{-1}) \, .
\end{align}                                                                   
The determinant is calculated by following the same procedure in \cite{Mei},
\begin{align}                                                                 
    {\rm det}^{1/2} (G^{-1})  
    &= \bigl[ p_0^2 - E_{\Delta}^{+\,2} \bigr]^2
       \bigl[ p_0^2 - E_{\Delta}^{-\,2} \bigr]^2
       \bigl[ p_0^2 - E^{+\,2} \bigr]
       \bigl[ p_0^2 - E^{-\,2} \bigr],
\end{align}                                                                   
where $p_0 = i(2n+1)\pi T ,\, (n=\cdots,-2,-1,0,1,2,\cdots)$ and
\begin{align}                                                                 
 & E_{\Delta}^\pm{}^2 \equiv (E \pm \mu)^2 + |\Delta|^2 \,, \,\,
   E^\pm \equiv E \pm \mu\,,  E \equiv \sqrt{p_1^{\,\, 2} + \sigma^2} \,.
\label{energy}
\end{align}                                                                   
Then, we obtain
\begin{align}                                                                 
     \Omega & (\sigma,|\Delta|) = 
      \frac{\sigma^2}{4G_S} + \frac{|\Delta|^2}{4G_D} \nonumber \\
     & {} - 2T \sum_{\pm} \sum_n \int \!\! \frac{dp_1}{2\pi}
     \biggl[
     \ln[\beta^2 (p_0^2 - E^{\pm}{}^2)] 
     + 2 \ln[\beta^2 (p_0^2 - E_{\Delta}^{\pm}{}^2)]      
     \biggr].
\label{Thermody}
\end{align}                                                                   

The frequency summation is performed through using the standard method
\cite{LeBellac},
\begin{align}                                                                 
 \sum_n \ln [\beta^2 (p_0^2 - E^2)]= \beta [ E + 2 T \ln( 1+e^{-\beta E} ) ].
\end{align}                                                                   
Thus we finally arrive at
\begin{align}                                                                 
     \Omega(\sigma,|\Delta|) &= \Omega_{0}(\sigma,|\Delta|)
            + \Omega_{T}(\sigma,|\Delta|), \label{Thermo3D} \\
     \Omega_0  (\sigma,|\Delta|) &= 
      \frac{\sigma^2}{4G_S} + \frac{|\Delta|^2}{4G_D}
      -2  \int \!\! \frac{d p_1}{2\pi}
     \bigl[
     E + E_{\Delta}^+ + E_{\Delta}^-
     \bigr], \label{Thermo2D0} \\
     \Omega_T  (\sigma,|\Delta|) &= 
      -2 T \sum_{\pm} \int \!\! \frac{d p_1}{2\pi}
     \biggl[
        \ln (1+e^{-\beta E^{\pm}}) 
        + 2 \ln (1+e^{-\beta E_{\Delta}^{\pm}})
     \biggr]. \label{Thermo3DT}
\end{align}                                                                   
$\Omega_0$ is $T$ independent contribution, which is ultraviolet
divergent, while $T$ dependent part $\Omega_T$ is finite.
For the purpose of later use, we rewrite $\Omega_0$ in the
integral form in the 2D Euclidean momentum space,
\begin{align}
     \Omega_0 (\sigma,|\Delta|) = 
      \frac{\sigma^2}{4G_S} + \frac{|\Delta|^2}{4G_D} 
      - \sum_{\pm} \int \!\! \frac{d^2 p_E}{(2\pi)^2}
     \biggl[
     \ln \Bigl( \frac{p_{E0}^2 + E^{\pm}{}^2}{p_E^2} \Bigr) 
     + 2 \ln \Bigl( \frac{p_{E0}^2 + E_{\Delta}^{\pm}{}^2}{p_E^2} \Bigr)
     \biggr].
\label{thermo02d}
\end{align}
The $p_E^2$ terms in the denominators in the arguments of logarithms
are inserted so as to drop an irrelevant infinite constant.

\subsection{Renormalized thermodynamic potential}
To eliminate the divergence, we perform the renormalization
by introducing the counter Lagrangian 
density\cite{Kohyama}, $\mathcal{L}_C = -Z_S \sigma^2/2 -Z_D|\Delta|^2$.
The renormalization factors $Z_S$ and $Z_D$ become
\begin{align}                                                                 
 Z_S =  \frac{6}{\pi\epsilon} + \frac{6}{\pi}\ln\alpha^2, \nonumber \\
 Z_D =  \frac{2}{\pi\epsilon} + \frac{2}{\pi}\ln\alpha^2,
\end{align}                                                                   
where $\epsilon\, (\equiv 2 - D)$ is the dimensional regularization parameter and 
$\alpha$ is an arbitrary renormalization scale.

Let us introduce the following parameters
\begin{align}                                                                 
  \sigma_0 & \equiv |\alpha| e^{-\pi/(12G_S)}, \label{sigma0} \\
  \Delta_0 & \equiv |\alpha| e^{-\pi/(8G_D)}, \label{Delta0}
\end{align}                                                                   
where $\sigma_0$ ($\Delta_0$) is the $q\bar{q}$ ($qq$) condensate at $T=0=\mu$
in the model in which the $qq$ ($q\bar{q}$) condensate is absent.
By using the above parameters, we finally obtain the renormalized thermodynamic
potential:
\begin{align}                                                                 
    \Omega_{r}&(\sigma,|\Delta|) = \Omega_{0r}(\sigma,|\Delta|)
      + \Omega_T(\sigma,|\Delta|)  \label{re_thermo}\,, \\
     \Omega_{0r} & (\sigma,|\Delta|) = 
      - \Bigl( \frac{3\pi}{2} \ln \sigma_0^2 \Bigr) \sigma^2
      - \Bigr( \pi \ln \Delta_0^2 \Bigr) |\Delta|^2 \nonumber \\
     & {} - \sum_{\pm} \int \!\! \frac{d^2 p_E}{(2\pi)^2}
     \biggl[
     \ln \Bigl( \frac{p_{E0}^2 + E^{\pm}{}^2}{p_E^2} \Bigr) 
     + 2 \ln \Bigl( \frac{p_{E0}^2 + E_{\Delta}^{\pm}{}^2}{p_E^2} \Bigr) 
      -\frac{3}{p_E^2} \sigma^2 -\frac{2}{p_E^2} |\Delta|^2
     \biggr].
\label{rezero}
\end{align}                                                                   
Note that the thermodynamic potential $\Omega_r$ has two
free parameters ($\sigma_0$, $\Delta_0$).
As in \cite{Kohyama}, we take $\sigma_0$
to be the scale of the theory ($\sigma_0 \geq 0$), and
choose the ratio $\Delta_0/\sigma_0$ as a parameter.

\section{The chiral and diquark condensates}
In this section, we show the numerical results for the chiral and diquark condensates
at zero and finite temperatures. Through minimizing the thermodynamic potential,
we study the $q\bar{q}$ and $qq$ condensates.

\begin{figure*}
 \begin{center}
 \includegraphics[width=14cm]{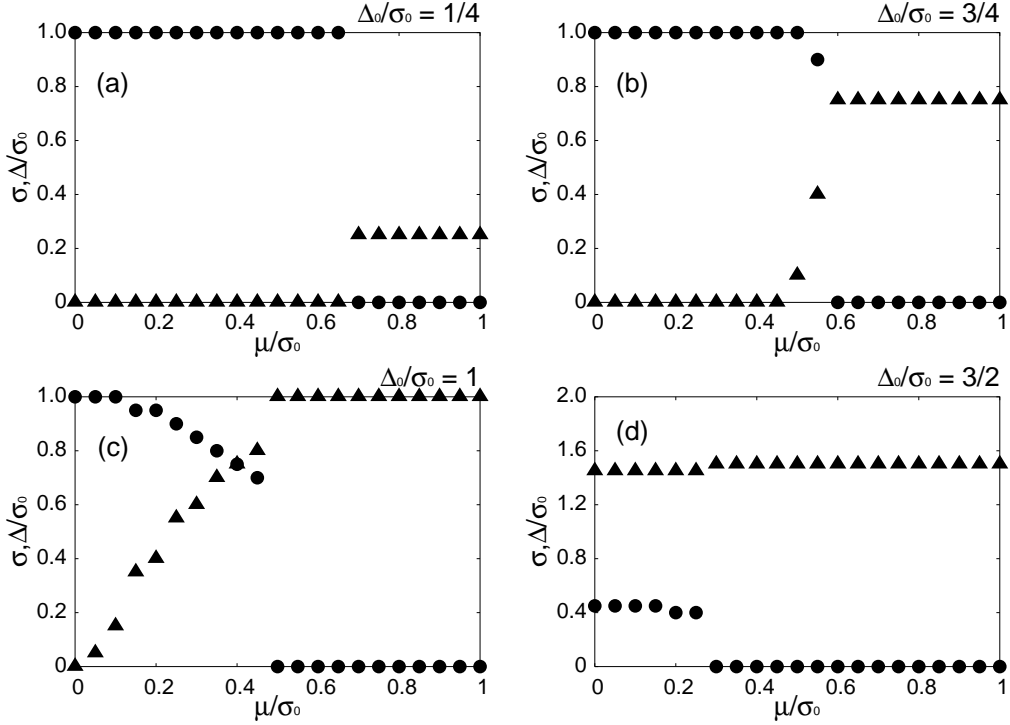}
 \end{center}
 \caption{\label{fig:condensates}The chiral(circles) and diquark(triangles)
  condensates at $T=0$.}
\end{figure*}
In Fig.~\ref{fig:condensates}, we display the results for the $q\bar{q}$ and $qq$
condensates at $T=0$. For $\Delta_0/\sigma_0=1/4$ (panel (a)), we see that the $q\bar{q}$
condensate is realized for $\mu=0 \sim 0.65 \sigma_0$ and there dose not arise
the $qq$ condensate. At $\mu=0.7 \sigma_0$, the $q\bar{q}$ condensate
disappears and the $qq$ condensate occurs, which indicates the phase transition from
$q\bar{q}$ condensed phase to the $qq$ condensed phase. Similar results are obtained for
$\Delta_0/\sigma_0=3/4,\,1$ where the $q\bar{q}$ condensate is dominant at low
$\mu$ and the $qq$ condensate dominates at high $\mu$. However
in the panel (d) ($\Delta_0/\sigma_0=3/2$), the $qq$ condensate is dominant for whole
$\mu$, and the $q\bar{q}$ condensate is small compare to the $qq$ condensate.
Thus the behaviors of the $q\bar{q}$ and $qq$ condensates are sensitive to the
value of the ratio $\Delta_0/\sigma_0$.

\begin{figure*}
\begin{center}
\includegraphics[width=14cm]{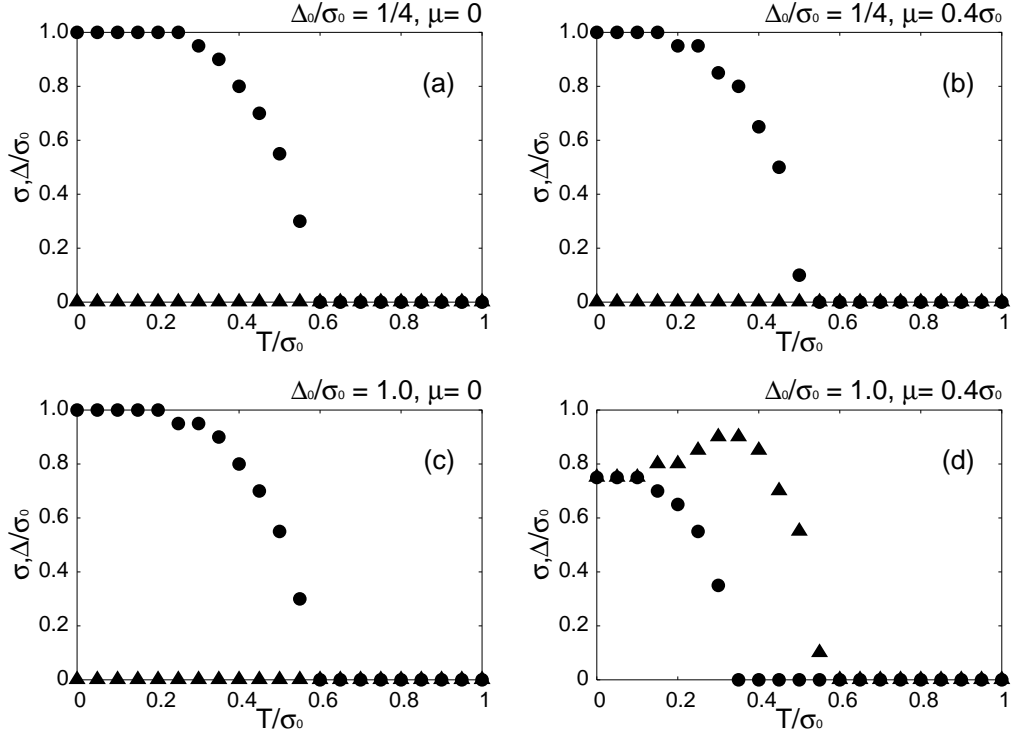}
\end{center}
\caption{\label{fig:Tconden}The $q\bar{q}$(circles) and $qq$(triangles) condensates at finite
    temperature.}
\end{figure*}
In Fig.~\ref{fig:Tconden}, we plot the results at finite temperature for 
$\Delta_0/\sigma_0=1/4,\,1$ and $\mu = 0,\, 0.4\sigma_0$. For
($\Delta_0/\sigma_0$, $\mu$) $=$ ($1/4$, $0$), the $q\bar{q}$ 
condensate starts to decrease at $T = 0.25 \sigma_0$ and completely disappears at 
$T = 0.6 \sigma_0$. For the cases $(\Delta_0/\sigma_0,\,\mu) = (1/4,\,0.4\sigma_0)$
and $(1,\, 0)$, we observe the similar behavior: The $q\bar{q}$ condensate decreases when
$T$ increases and the $qq$ condensate does not arise. In the panel (d)
($\Delta_0/\sigma_0$, $\mu$) $=$
($1$, $0.4\sigma_0$), the $qq$ condensate exists and it decreases with increasing
temperature.
Thus, the $q\bar{q}$ and $qq$ condensates decrease when $T$ becomes larger,
which is the reasonable result because the theory is asymptotically free.
We have carried out the numerical calculation for the other values of
$\Delta_0/\sigma_0$, and found that the behaviors do not alter qualitatively.

It is to be noted that the $q\bar{q}$ condensate reduces to zero discontinuously
at $T=0$ (see Fig.~\ref{fig:condensates}). While it decreases continuously
at finite temperature for some parameter region.
These are the signals of the first order
and second order phase transitions. The above mentioned results indicate that the
critical point from first order phase transition to second order one appears in
the phase diagram, which we will discuss in more detail in the next section.

\section{The phase diagram}
\begin{figure*}
\begin{center}
\includegraphics[width=14cm]{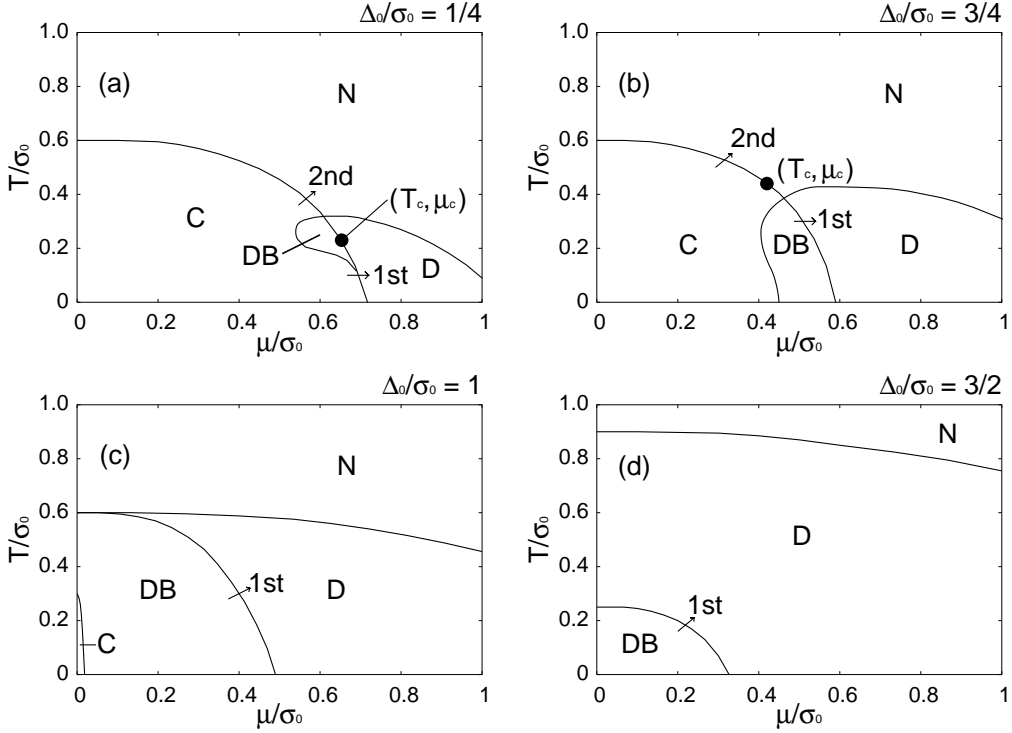}
\end{center}
\caption{\label{fig:phd}The phase diagrams.}
\end{figure*}
Fig.~\ref{fig:phd} displays the phase diagrams for the cases 
$\Delta_0/\sigma_0 = 1/4$, $3/4$, $1$, $3/2$. As seen in the diagrams, there appear
the following phases:
\begin{quote}
 C: Chiral condensed phase ($\sigma \neq 0$, $\Delta =0$)\\
 D: Diquark condensed phase($\sigma = 0$, $\Delta \neq 0$)\\
 DB: Double Broken phase ($\sigma \neq 0$, $\Delta \neq 0$)\\
 N: Normal phase ($\sigma = 0$, $\Delta = 0$)
\end{quote}
For small $\Delta_0/\sigma_0$, the C phase is realized at low $T$ and $\mu$,
the DB phase occurs at low $T$ and intermediate $\mu$ and the D phase
at low $T$ and high $\mu$. This phase structure bears resemblance to the QCD phase
diagram. However, when $\Delta_0/\sigma_0$ is large, the C phase no longer appears
and the DB phase exists at low $T$ and $\mu$, and the D phase does at low $T$ and
high $\mu$. Thus, as the ratio $\Delta_0/\sigma_0$ increases, the region of
the C phase becomes smaller and those of the DB phase and D phase enlarge.

The black circles ($T_{c}$, $\mu_{c}$) seen in Fig.~\ref{fig:phd}(a) and (b) are the critical points from
the first order phase transition to the second order one. When $T < T_c$, the
transition is of the first order and it is of the second order for $T > T_c$.
From the figure, we find that the critical point moves along the transition line
toward $\mu = 0$ as the ratio $\Delta_0/\sigma_0$ increases. The critical
point no longer appears at $\Delta_0/\sigma_0 = 1$, and the transition
from the DB phase to D phase is of the first order.
The other phase transitions, D $\rightarrow$ N and C $\rightarrow$ DB, are always
of the second order.

\section{Concluding remarks}
Through studying the $q\bar{q}$ and $qq$ condensates at finite temperature and
density, we have obtained phase diagrams. When the ratio $\Delta_0/\sigma_0$
is small, there appears the chiral condensed (C)
phase at low $T$ and $\mu$, the diquark condensed (D) phase at low $T$ and high
$\mu$, and the double broken (DB) phase at low $T$ and intermediate $\mu$.
On the other hand, for large $\Delta_0/\sigma_0$, the DB phase is realized at
low $T$ and $\mu$ and the D phase at low $T$ and high $\mu$. Thus the phase
structure drastically changes with respect to the ratio $\Delta_0/\sigma_0$.
Such a behavior of the phase diagrams shows close similarity with the ones in the
3D GN model.

The direct comparison of the present model and the 4D NJL model is difficult
since the free parameter of these models are different. In the 4D NJL model, the
parameter is the ratio of two coupling constants $G_D/G_S$\cite{Mei}, and,
in the present model, the parameter is the ratio
$\Delta_0/\sigma_0$. However $\Delta_0$ is related to $G_D$ through
Eq.(\ref{Delta0}), and it increases as $G_D$ increases. This means
that $\Delta_0/\sigma_0$ increases as $G_D$ becomes larger. Then, with increasing
$G_D$, the phase diagram changes from Fig.~\ref{fig:phd}(a) to (d). After all this,
we see that the phase structure bears resemblance to the 4D NJL case.

Having compared the 2D GN model, 3D GN model and 4D NJL model, we see that the
similar phase structure is realized in all models. This indicates that the phase
structure of $q\bar{q}$ and $qq$ condensates in the NJL-type model
does not change appreciably in 2D, 3D and 4D.

\begin{acknowledgments}
 I would like to thank to A. Niegawa for numerous discussions and reading of the
 manuscript. I also thank to T. Inagaki for useful discussions.
\end{acknowledgments}


\end{document}